# EINSTEIN"S COEFFICIENTS AND THE NATURE OF THERMAL BLACKBODY RADIATION

Fedor V.Prigara

*Institute of Microelectronics and Informatics, Russian Academy of Sciences,*

*21 Universitetskaya, 150007 Yaroslavl, Russia*

We show that thermal radio emission has an induced character and argue that thermal blackbody radiation in other spectral ranges also has an induced origin. A new theory of thermal radio emission of non-uniform gas basing on the induced origin of emission and its astrophysical applications are considered. The nature of emission from various astrophysical objects is discussed.

PACS numbers: 95.30.Gv, 42.52.+x.

The analysis of observational data on thermal radio emission from various astrophysical objects leads to the condition of emission implying the induced emission of radiation [5]. Here we show that the stimulated character of thermal radio emission follows from the relations between Einstein's coefficients for spontaneous and induced emission of radiation.

Consider two level system (atom or molecule) with energy levels $E_1$ and $E_2 > E_1$ which is in equilibrium with thermal blackbody radiation. We denote as $\nu_s = A_{21}$ the number of transitions from the upper energy level to the lower one per unit time caused by a spontaneous emission of radiation with the frequency $\omega = (E_2 - E_1)/\hbar$, where $\hbar$ is the Planck constant. The number of transitions from the energy level $E_2$ to the energy level $E_1$ per unit time caused by a stimulated emission of radiation may be written in the form

$$\nu_i = B_{21} B_\nu \Delta\nu, \qquad (1)$$

where

$$B_\nu = \hbar\omega^3/(2\pi^2 c^2)(\exp(\hbar\omega/kT) - 1) \qquad (2)$$

is a blackbody emissivity (Planck's function), $T$ is the temperature, $k$ is the Boltzmann constant, $c$ is the speed of light, and $\Delta\nu$ is the line width.

The number of transitions from the lower level to the upper one per unit time may be written in the form

$$\nu_{12} = B_{12} B_\nu \Delta\nu, \qquad (3)$$

Here $A_{21}$, $B_{21}$ and $B_{12}$ are the coefficients introduced by Einstein [1,2], the coefficients $B_{12}$ and $B_{21}$ being modified to account for the line width $\Delta\nu$.

We denote as $N_1$ and $N_2$ the number of atoms occupying the energy levels $E_1$ and $E_2$, respectively. The levels $E_1$ and $E_2$ are suggested for simplicity to be non-degenerated.

In the equilibrium state the full number of transitions from the lower level to the upper one is equal to the number of reverse transitions:

$$N_1 \nu_{12} = N_1 B_{12} B_\nu \Delta\nu = N_2 \nu_{21} = N_2 (A_{21} + B_{21} B_\nu \Delta\nu). \qquad (4)$$

From the last equation we obtain

$$N_2 / N_1 = B_{12} B_\nu \Delta\nu / (A_{21} + B_{21} B_\nu \Delta\nu). \qquad (5)$$

The line width $\Delta\nu$ is suggested to be equal to the natural line width [3,4]:

$$\Delta\nu = A_{21} + B_{21} B_\nu \Delta\nu. \qquad (6)$$

It follows from the last equation that

$$\Delta\nu = A_{21} / (1 - B_{21} B_\nu). \qquad (7)$$

Substituting this expression in the equation (5), we obtain

$$N_2 / N_1 = B_{12} B_\nu. \qquad (8)$$

In the limit of higher temperatures $T \to \infty$, corresponding to the range of frequencies $\hbar\omega < kT$, the function $B_\nu(T)$ is given by the Rayleigh-Jeans formula

$$B_\nu = 2kT\nu^2 / c^2, \qquad (9)$$

where $\nu$ is the frequency of radiation, $\nu = \omega / 2\pi$.

Since $B_\nu \to \infty$ when $T \to \infty$, it follows from the equation (7) that the coefficient $B_{21}$ is depending on the temperature $T$ in such a manner that $B_{21} B_\nu < 1$. It is clear that $B_{21} B_\nu \to 1$ when $T \to \infty$. The ratio of frequencies of transitions caused by spontaneous and induced emission of radiation is given by the expression

$$\nu_s / \nu_i = A_{21} / (B_{21} B_\nu \Delta\nu) = (1 - B_{21} B_\nu) / (B_{21} B_\nu). \qquad (10)$$

This ratio is approaching zero when $T \to \infty$, since $B_{21} B_\nu \to 1$. It means that in the range of frequencies $\hbar\omega < kT$ thermal radiation is produced by the stimulated emission, whereas the contribution of a spontaneous emission may be neglected.

The quantum of radio emission has the energy $\hbar\omega < 10^{-3} eV$. In the same time the temperature of radio emitting gas in the case of astrophysical objects usually exceeds $10^2$ K, so the condition $\hbar\omega < kT$ is valid. Thus thermal radio emission from gas nebulae and other astrophysical objects has the induced origin.

Thermal radio emission of non-uniform gas is produced by the ensemble of individual emitters. Each of these emitters is a molecular resonator the size of which has an order of magnitude of mean free path of photons

$$l = \frac{1}{n\sigma} \qquad (11),$$

where $n$ is the number density of particles of gas, and $\sigma$ is the absorption cross-section. The absorption cross-section has an order of magnitude of atomic cross-section $\sigma=10^{-15}$ cm$^2$. Because of a stimulated emission, the attenuation of radio emission will be sufficiently small, although the value of absorption cross-section is relatively great. In the range of frequencies $\hbar\omega < kT$ $B_{12}B_\nu = N_2/N_1 \approx 1$ and therefore $B_{12}B_\nu \approx B_{21}B_\nu$. It means that the number of transitions from the lower energy level to the upper one caused by the absorption of photon will be approximately equal to the number of reverse transitions from the upper level to the lower one caused by the stimulated emission of quantum of the same character as the original one.

Note that owing to the induced absorption of light described above Kirchhoff's law [1] is not valid in the range of frequencies $\hbar\omega < kT$, where the induced emission of radiation dominates and $B_{12}B_\nu \approx B_{21}B_\nu$.

The emission of each molecular resonator is coherent, with the wavelength

$$\lambda = l, \qquad (12)$$

and thermal radio emission of gaseous layer is incoherent sum of radiation produced by individual emitters.

The condition (12) implies that the radiation with the wavelength $\lambda$ is produced by the gaseous layer with the definite number density of particles $n$.

In the gaseous disk model, describing radio emitting gas nebulae [5], the number density of particles decreases reciprocally with respect to the distance $r$ from the energy center

$$n \propto r^{-1}. \qquad (13)$$

Together with the condition of emission (12) the last equation leads to the wavelength dependence of radio source size:

$$r_\lambda \propto \lambda . \qquad (14)$$

The relation (14) is indeed observed for sufficiently extended radio sources [6,7,8]. For example, the size of radio core of galaxy M31 is 3.5 arcmin at the frequency 408 MHz and 1 arcmin at the frequency 1407 MHz [8].

In the case of some compact radio sources instead of the relation (14) the relation

$$r_\lambda \propto \lambda^2 \qquad (15)$$

is observed [9]. This relation may be explained by the effect of a gravitational field on the distribution of gas density which changes the equation (13) for the equation

$$n \propto r^{-1/2}. \qquad (16)$$

The spectral density of flux from an extended radio source is given by the formula

$$F_\nu = \frac{1}{a^2} \int_0^{r_\lambda} B_\nu(T) \times 2\pi r \, dr, \qquad (17)$$

where $a$ is a distance from radio source to the detector of radiation.

The extended radio sources may be divided in two classes. Type 1 radio sources are characterized by a stationary convection in the gaseous disk with an approximately uniform distribution of the temperature $T \approx const$ giving the spectrum

$$F_\nu \approx const. \qquad (18)$$

Type 2 radio sources are characterized by outflows of gas with an approximately uniform distribution of gas pressure $P = nkT \approx const$. In this case the equation (13) gives

$$T \propto r, \qquad (19)$$

so the radio spectrum, according to the equation (17), has the form

$$F_\nu \propto \nu^{-1}. \qquad (20)$$

Both classes include numerous galactic and extragalactic objects. In particular, edge-brightened supernova remnants [10] belong to the type 2 radio sources in accordance with the relation (19), whereas center-brightened supernova remnants belong to the type 1 radio sources. Steep spectrum radio quasars [15] belong to type 2 radio sources, and flat spectrum radio quasars [15] belong to type 1 radio sources.

There is a direct observational evidence that the outflows of gas have radio spectra describing by the equation (20). This evidence was obtained for the bright radio condensations emerging in opposite directions from a compact core of the radio counterpart of GRS1915+105 [11].

Note that the synchrotron interpretation of radio emission from various astrophysical objects [16] encounters the essential difficulties. The synchrotron theory is unable to explain the correlation between the spectral index and the radial distribution of brightness for supernova remnants [10], as well as the wavelength dependence of radio source size , since a spectral index according to the synchrotron theory has a local origin [16]. The constancy of the spectral index for a jet [17] also cannot be explained by the synchrotron theory predicting the evolution of a radio spectrum to the steeper one.

Planetary atmospheres are similar to type 1 radio sources , since the temperature of radio emitting atmosphere layers is approximately constant , whereas the gas pressure is exponentially decreasing with the increase of height. Since the total thickness of radio emitting layers is small with respect to the radius of a planet , the spectrum of thermal radio emission of planetary atmosphere is similar to the radio spectra of planetary nebulae in the long wavelengths range [12]. Notice that at the higher frequencies radio emission of planetary nebulae is described by the law (18).

The induced origin of thermal radio emission is consistent with the existence of maser sources associated with gas nebulae and nuclei of galaxies [13,14].

The higher brightness temperatures of compact extragalactic radio sources [6] are explained by maser amplification of thermal radio emission. This conclusion is supported by the constancy of the brightness temperature of the radio core of quasar NRAO 530 in the wavelength range of 0.3 cm to 2 cm , the flux density obeying the law (18) in the wavelength range of 0.35 cm to 6.25 cm [6]. It is quite possible that the higher brightness temperatures of radio pulsars are also connected with maser amplification of thermal radio emission.

It is worthwhile to notice that the original Einstein's theory [1] which does not take into account a line width also predicts the induced character of thermal radio emission. However the account for the natural line width essentially affects the form of relation (8).

In the limit of $T \to 0$, corresponding to the range of frequencies $\hbar\omega > kT$, the function $B_\nu(T)$ is given by the formula

$$B_\nu = (\hbar\omega^3 / 2\pi^2 c^2)\exp(-\hbar\omega / kT) \qquad (21)$$

and the relation (8) gives Boltzmann's law

$$(N_2 / N_1) = \exp(-\hbar\omega / kT) \qquad (22)$$

if we take

$$B_{12} = 2\pi^2 c^2 / \hbar\omega^3. \qquad (23)$$

If we suggest in addition that $B_{12} = B_{21}$, then the line width will be given by the formula

$$\Delta\nu = A_{21} / (1 - \exp(-\hbar\omega / kT)). \qquad (24)$$

The ratio of frequencies of transitions caused by induced and spontaneous emission of radiation respectively is given by the expression

$$\nu_i / \nu_s = B_{21} B_\nu / (1 - B_{21} B_\nu) \approx \exp(-\hbar\omega / kT). \qquad (25)$$

In this range of frequencies the contribution of spontaneous emission dominates.

Notice that the relation $B_{12} = B_{21}$ is no more than hypothesis which cannot be justified by thermodynamical approach . In fact this hypothesis might not be valid and then the ratio $\nu_i / \nu_s$ may exceed the unit. It is quite possible since the coefficients $B_{12}$, and $B_{21}$ are depending on the temperature.

The strong argument in favor of induced character of thermal blackbody emission is that the whole range of spectrum is described by a single Planck's function. So if thermal radio emission is stimulated one then thermal emission in other spectral ranges also should have the induced character.

Since the existence of maser sources is closely related with the induced character of thermal radio emission , one may expect that laser sources in other spectral ranges also can exist.

One possible example of laser source is gamma-laser at the Galactic Center emitting in the line 0.511 MeV [18].

Another effect presumably related with laser sources is the large fluctuations in widths of solar EUV lines [19]. Note that the Sun as an aggregate of small cells with random phases [19] may be compared with an aggregate of molecular resonators described above.

Rapid variations of [OIII] line profiles in a Seyfert 2 galaxy [17] is similar effect. The polarization of optical emission lines in the spectra of radio galaxies [7] and of rapidly varying continuum in the spectra of BL Lacertae objects [15] also may be explained by a laser mechanism of emission.

Molecular resonator can emit second and higher harmonics. Such thermal harmonics are likely to be observed in infrared spectrum of planetary nebula NGC 7027 [12]. There is a clear indication of second thermal harmonic in the optical spectrum of the Sun. This effect may determine partially or in whole the deviation of the spectral energy distribution of the Sun from Planck's function.

Thus it follows from the relations between Einstein's coefficients that thermal radio emission has a stimulated character, and it is quite possible that thermal blackbody radiation in other spectral ranges also has an induced origin. The induced origin of thermal radiation leads to a new theory of thermal radio emission of non-uniform gas having broad astrophysical applications and suggests the existence of laser sources in optical and other spectral ranges.

The author is grateful to V.V.Ovcharov and A.V.Postnikov for useful discussions.

----------------------------------------------------------------------------------------------------------